\documentclass[conference]{IEEEtran}

\ifCLASSINFOpdf
 \else
\fi
\usepackage[tight,footnotesize]{subfigure}
\usepackage[font=footnotesize]{subfig}
\usepackage{fixltx2e}
\usepackage{stfloats}
\usepackage{url}

\usepackage{enumitem}

\usepackage[pdftex]{graphicx}

\begin{document}

\title{Live Inspection of Spreadsheets}

\author{\IEEEauthorblockN{Daniel Kulesz\IEEEauthorrefmark{1},
Fabian Toth\IEEEauthorrefmark{2} and
Fabian Beck\IEEEauthorrefmark{3}}
University of Stuttgart
\IEEEauthorblockA{\IEEEauthorrefmark{1}daniel.kulesz@informatik.uni-stuttgart.de, \IEEEauthorrefmark{2}tothfn@studi.informatik.uni-stuttgart.de, \IEEEauthorrefmark{3}fabian.beck@visus.uni-stuttgart.de}
}

\maketitle

\begin{abstract}



Existing approaches for detecting anomalies in spreadsheets can help to discover faults but they are often applied too late in the spreadsheet lifecycle. By contrast, our approach detects anomalies immediately whenever users change their spreadsheets. This live inspection approach has been implemented as part of the Spreadsheet Inspection Framework, enabling the tool to visually report findings without disturbing the users' workflow. An advanced list representation allows users to keep track of the latest findings, prioritize open problems, and check progress on solving the issues. Results from a first user study indicate that users find the approach useful.

\end{abstract}

\IEEEpeerreviewmaketitle

\section{Introduction}
Faults in spreadsheets are common and can cause severe damage \cite{powell2008critical}. In recent years, several tool-based approaches have been developed for automatically detecting anomalies in spreadsheets \cite{jannach2014avoiding}, arguing that anomalies are dependable indicators for possible faults in spreadsheets. Today, a number of anomaly detection tools aims for a tight integration into spreadsheet environments like Microsoft Excel. One key benefit of tightly integrated tools is their ability to communicate findings in the same environment users work with their spreadsheets. Findings vary by detection approach and can be, e.g. smelly formulas, failed test cases, or violated constraints.


Existing spreadsheet anomaly detection tools still suffer from a major drawback: Since their scans have to be triggered manually, it cannot be guaranteed that users execute them regularly---if at all. The more actions users take and the more time passes between scans, the more difficult it gets for users to identify the actions responsible for the reported findings. Also, the spreadsheet could contain findings already created by previous users, and distinguishing between new and old findings puts an additional mental load onto users.

\begin{figure}[!h]
\centering
\includegraphics[width=3.1in]{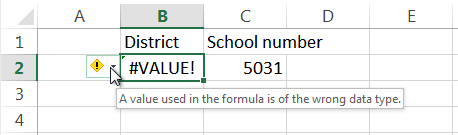}

\caption{Anomaly reported by Microsoft Excel 2013}
\label{fig_valueerror}
\end{figure}

At first glance, checking spreadsheets for anomalies automatically in the background seems to be the trivial solution to address the described issues. In fact, recent versions of Microsoft Excel already issue live inspection techniques by providing warning icons for built-in error types and report them as warning diamonds next to the affected cells (Figure \ref{fig_valueerror}). However, adopting a similar approach for more complex anomalies involves several challenges. In the rest of this position paper, we describe these challenges, our attempts to overcome them, and first experiences with the resulting solution.

\section{Live inspection challenges}

We identified the following challenges for a live inspection approach of spreadsheets:

\begin{enumerate}[leftmargin=*, label= Ch\arabic*:]
  \item Avoid disruption: Users primarily want to work with their spreadsheet and not study findings. Thus, users must be notified in a low-disrubtive manner.
  \item Support workflow: Users should be free to choose when to deal with findings and supported in their workflow addressing the findings.
  \item Motivation: Users should be motivated to deal with reported findings.
\item Recent first: Findings caused by recent user actions should be reported in a more prominent fashion than older findings.
\item Provide overview: Users need an overview of all currently open findings and support to group them.
  
\end{enumerate}

\section{Live Inspection Approach}

\begin{figure}[!t]
\centering
\includegraphics[width=3.5in]{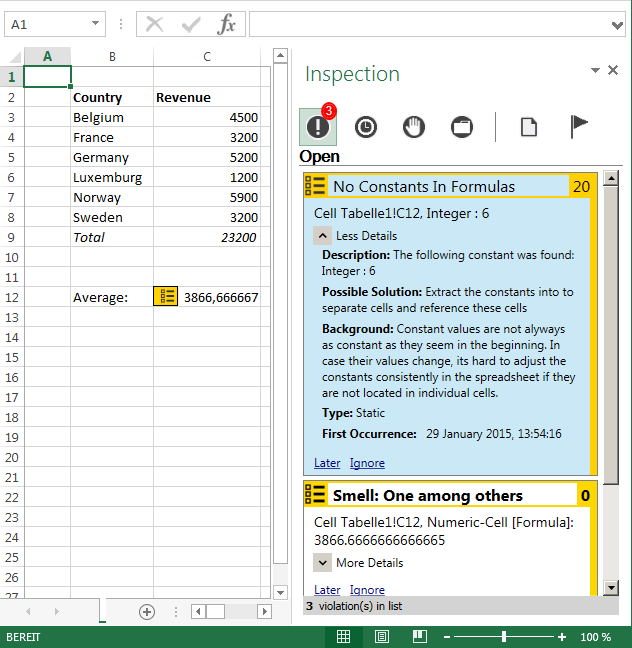}

\caption{Extended inspection pane of SIF}
\label{fig_inspectionpane}
\end{figure}

In previous work \cite{kulesz2014integrating}, we developed an open-source tool named Spreadsheet Inspection Framework (SIF)\footnote{http://spreadsheet-inspection-framework.github.io}. It allows users to scan spreadsheets for anomalies using a number of automated and partly automated detection techniques. SIF visualizes findings in a side pane and with in-spreadsheet marker icons (example in Figure \ref{fig_inspectionpane}, cell C12). We extend this tool by a live inspection mechanism.

The live inspection mechanism borrows ideas from the task metaphor that many e-mail clients employ to empower users to process and categorize incoming e-mails \cite{szostek2011dealing}. Users typically do not sit and wait for new e-mails to arrive but do other work. Similarly, spreadsheet users are trying to solve a task with their spreadsheet when being informed about a new finding they caused. Even if the findings are a direct cause of the users' actions, the situation may be comparable in terms of the amount of attention users are willing to invest.

\begin{figure}[!t]
\centering
\includegraphics[width=2.55in]{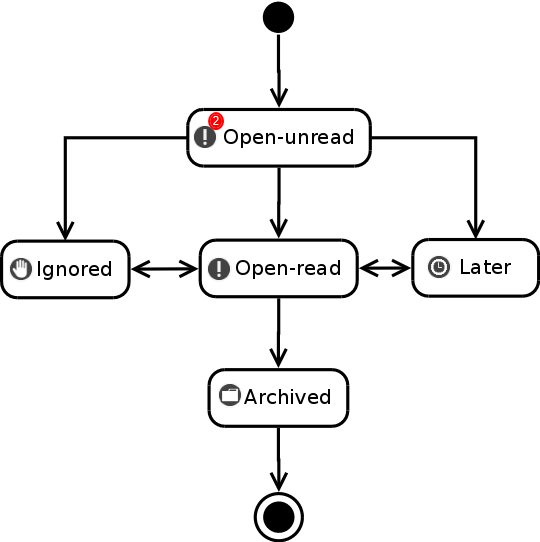}

\caption{State diagram for finding states}
\label{fig_statechart}
\end{figure}

To adopt the metaphor, we divided the formerly flat list (Ch5) of findings (anomalies) into four categories represented by the tabs shown in Figure \ref{fig_inspectionpane}. They are named, from left to right: `Open', `Later' (postponed), `Ignored' and `Archive' (solved). Figure \ref{fig_statechart} illustrates the states between which findings can travel.

A newly detected finding is added to the list of findings as 'Open'. Open findings have a state of being `read' or `unread'. The eye-catching red bubble next to the `Open' section shows the number of unread findings (Ch1 Ch3, Ch5). Recent findings are placed on top (Ch4). The tabs `Later' and `Ignore' do not employ this notification mechanism because findings become part of these categories only when users explicitly move them there (Ch2)---by definition, all findings of these categories are read.

Findings that were raised but are meanwhile solved move to the `Archive' without triggering a notification. The archive is a neutral category. In a preliminary version we experimented with a reward mechanism that highlighted solved findings and counted them using a green notification bubble (Ch3), but since many findings can be `solved' by simply deleting the causative cells, this counter could introduce false incentives.

On the technical side, we implemented a diff mechanism which allows us to distinguish new from existing findings. Additionally, we set a trigger that automatically issues a scan whenever a user's action triggers a recalculation of the spreadsheet. An extended preferences dialog allows users to choose which inspection rules shall be included in the automatic scan on a per-spreadsheet basis. 

\section{Evaluation}

We did a first user study with one pilot followed by five participants (male engineering students aged between 18 and 24). The participants had to solve two tasks in a given spreadsheet that required extending and changing the business logic of its formulas. The spreadsheet was designed in such a way that findings were likely to be caused.

Three of the participants used the in-spreadsheet marker icons and directly tried to solve open issues as they appeared, but skipped findings they did not find trivial in the first place. They kept the inspection pane closed and did not open it until they finished their primary tasks. Then, they used the inspection pane to solve the remaining findings. The forth participant did not pay attention at all to the findings until he finished the primary task. In contrast, the fifth participant kept the inspection pane open all the time and paid more attention to the findings than to the actual task.

The first three subjects achieved a rapid learning effect: Once they understood that constants in formulas lead to findings, they changed their behavior and did not put constants into the next formulas they created. Overall, the participants rated the live inspection mechanism to be low distracting and generally acceptable.

\section{Future work}
 
Future work should provide an extended reward mechanism to motivate the users to solve open findings. When processing large spreadsheets or checking for more complex anomalies, users should be able to continue working while the scans are still running. Also, checks need to be executed incrementally and in order of priority, comparable to selecting and prioritizing regression tests in software engineering.










%


\bibliographystyle{IEEEtran}
\bibliography{bibliography}


\end{document}